\PassOptionsToPackage{unicode}{hyperref}
\PassOptionsToPackage{hyphens}{url}
\PassOptionsToPackage{dvipsnames,svgnames,x11names}{xcolor}
\documentclass[
  11pt,
]{article}
\usepackage{amsmath,amssymb}
\usepackage{setspace}
\usepackage{iftex}
\ifPDFTeX
  \usepackage[T1]{fontenc}
  \usepackage[utf8]{inputenc}
  \usepackage{textcomp} 
\else 
  \usepackage{unicode-math} 
  \defaultfontfeatures{Scale=MatchLowercase}
  \defaultfontfeatures[\rmfamily]{Ligatures=TeX,Scale=1}
\fi
\usepackage{lmodern}
\ifPDFTeX\else
\fi
\IfFileExists{upquote.sty}{\usepackage{upquote}}{}
\IfFileExists{microtype.sty}{
  \usepackage[]{microtype}
  \UseMicrotypeSet[protrusion]{basicmath} 
}{}
\usepackage{xcolor}
\usepackage[margin=1.2in]{geometry}
\usepackage{graphicx}
\makeatletter
\def\maxwidth{\ifdim\Gin@nat@width>\linewidth\linewidth\else\Gin@nat@width\fi}
\def\maxheight{\ifdim\Gin@nat@height>\textheight\textheight\else\Gin@nat@height\fi}
\makeatother
\setkeys{Gin}{width=\maxwidth,height=\maxheight,keepaspectratio}
\makeatletter
\def\fps@figure{htbp}
\makeatother
\NewDocumentCommand\citeproctext{}{}
\NewDocumentCommand\citeproc{mm}{%
  \begingroup\def\citeproctext{#2}\cite{#1}\endgroup}
\makeatletter
 \let\@cite@ofmt\@firstofone
 \def\@biblabel#1{}
 \def\@cite#1#2{{#1\if@tempswa , #2\fi}}
\makeatother
\newlength{\cslhangindent}
\newlength{\csllabelwidth}
\newenvironment{CSLReferences}[2] 
 {\begin{list}{}{%
  \setlength{\itemindent}{0pt}
  \setlength{\leftmargin}{0pt}
  \setlength{\parsep}{0pt}
  \ifodd #1
   \setlength{\leftmargin}{\cslhangindent}
   \setlength{\itemindent}{-1\cslhangindent}
  \fi
  \setlength{\itemsep}{#2\baselineskip}}}
 {\end{list}}
\usepackage{calc}

\usepackage[scaled=0.98]{inconsolata} \usepackage{float} \usepackage{caption} \usepackage{enumitem} \setlist[itemize,1]{label=$\circ$} \setlist[itemize,2]{label=$\diamond$} \edef\restoreparindent{\parindent=\the\parindent\relax} \usepackage[parfill]{parskip} \restoreparindent
\ifLuaTeX
  \usepackage{selnolig}  
\fi
\usepackage{bookmark}
\IfFileExists{xurl.sty}{\usepackage{xurl}}{} 
\hypersetup{
  pdftitle={A multilevel model with heterogeneous variances for snap timing in the National Football League},
  colorlinks=true,
  linkcolor={magenta},
  filecolor={Maroon},
  citecolor={magenta},
  urlcolor={magenta},
  pdfcreator={LaTeX via pandoc}}

\title{A multilevel model with heterogeneous variances for snap timing
in the National Football League}
\author{Quang Nguyen \qquad \qquad \qquad Ronald Yurko\\
\strut \\
Department of Statistics \& Data Science\\
Carnegie Mellon University\\
Pittsburgh, PA 15213\\
\texttt{\{quang,\,ryurko\}@stat.cmu.edu}}
\date{}

\begin{document}
\maketitle
\begin{abstract}
Player tracking data have provided great opportunities to generate novel
insights into understudied areas of American football, such as pre-snap
motion. Using a Bayesian multilevel model with heterogeneous variances,
we provide an assessment of NFL quarterbacks and their ability to
synchronize the timing of the ball snap with pre-snap movement from
their teammates. We focus on passing plays with receivers in motion at
the snap and running a route, and define the snap timing as the time
between the moment a receiver begins motioning and the ball snap event.
We assume a Gamma distribution for the play-level snap timing and model
the mean parameter with player and team random effects, along with
relevant fixed effects such as the motion type identified via a Gaussian
mixture model. Most importantly, we model the shape parameter with
quarterback random effects, which enables us to estimate the differences
in snap timing variability among NFL quarterbacks. We demonstrate that
higher variability in snap timing is beneficial for the passing game, as
it relates to facing less havoc created by the opposing defense. We also
obtain a quarterback leaderboard based on our snap timing variability
measure, and Patrick Mahomes stands out as the top player.\\
\strut 
\end{abstract}

\setstretch{1}
\setlength{\parskip}{0.25\baselineskip}

\section{Introduction}\label{sec:introduction}

Recent developments in sports analytics have been largely sparked by
player tracking data (\citeproc{ref-baumer2023big}{Baumer, Matthews, and
Nguyen 2023}; \citeproc{ref-kovalchik2023player}{Kovalchik 2023}). In
American football, the National Football League (NFL) has been
collecting tracking data since 2016 via the Next Gen Stats system, where
RFID tags are placed in the shoulder pads of players and inside the
football. This records the location and trajectory of every player on
the field within a play at a rate of 10 frames per second. The richness
and continuous nature of player tracking data provide substantial
opportunities to gain deeper insight into various aspects of football
that were not previously captured by discrete play-level data.

To make these fine-grained data accessible and promote public research,
the NFL organizes the annual Big Data Bowl competition, beginning in
2018. Each year, a sample of tracking data is publicly released to
accompany a competition theme on a particular area of football (e.g.,
secondary, special teams, linemen, etc.). Early peer-reviewed articles
that leverage Big Data Bowl data mainly provide assessment of offensive
production in American football. Deshpande and Evans
(\citeproc{ref-deshpande2020expected}{2020}) propose a framework for
determining the hypothetical completion probability for a passing play.
Chu et al. (\citeproc{ref-chu2020route}{2020}) detect the types of route
ran by NFL receivers via a model-based clustering approach. Yurko et al.
(\citeproc{ref-yurko2020going}{2020}) introduce a continuous-time play
valuation framework which features a model for the expected yards gained
at every moment within a play.

Apart from offensive performance evaluation, tracking data have also
driven considerable advancements in the assessment of NFL defenders.
This task was once very challenging, as position groups such as
defensive linemen simply lacked reliable metrics to grade their
performance prior to the availability of high-resolution data. Dutta,
Yurko, and Ventura (\citeproc{ref-dutta2020unsupervised}{2020}) use a
mixture model to provide labels for the types of pass coverage by
defensive backs. Nguyen, Yurko, and Matthews
(\citeproc{ref-nguyen2024here}{2024}) present a novel metric for
quantifying defensive pressure by pass rushers at every frame within a
play. Yurko, Nguyen, and Pelechrinis (\citeproc{ref-yurko2024nfl}{2024})
evaluate defensive pass coverage by comparing the defender positioning
at the moment of catch to a baseline hypothetical player. Nguyen et al.
(\citeproc{ref-nguyen2025fractional}{2025}) introduce a framework for
within-play tackling evaluation, overcoming the limitations of
traditional box-score tackles statistics.

Similar to defensive evaluation, it is possible to investigate other
areas of football that were relatively understudied in the past with
tracking data. The 2025 edition of the NFL Big Data Bowl focuses on
pre-snap motion, which has become an important strategic aspect in
American football. Generally, pre-snap motion refers to the movement of
a player before the football is snapped to start the play. By employing
motion, a team's ultimate goal is to distract its opponent and gain an
advantage before the play even begins. For the first time ever, we have
access to on-field player locations for the full play---not just after
the snap but also before the snap---via the Big Data Bowl 2025 tracking
data sample. This opens up an opportunity to leverage pre-snap
information to better understand team and player behavior during a play
post-snap.

\begin{figure}[ptb]

{\centering \includegraphics[width=1.01\linewidth]{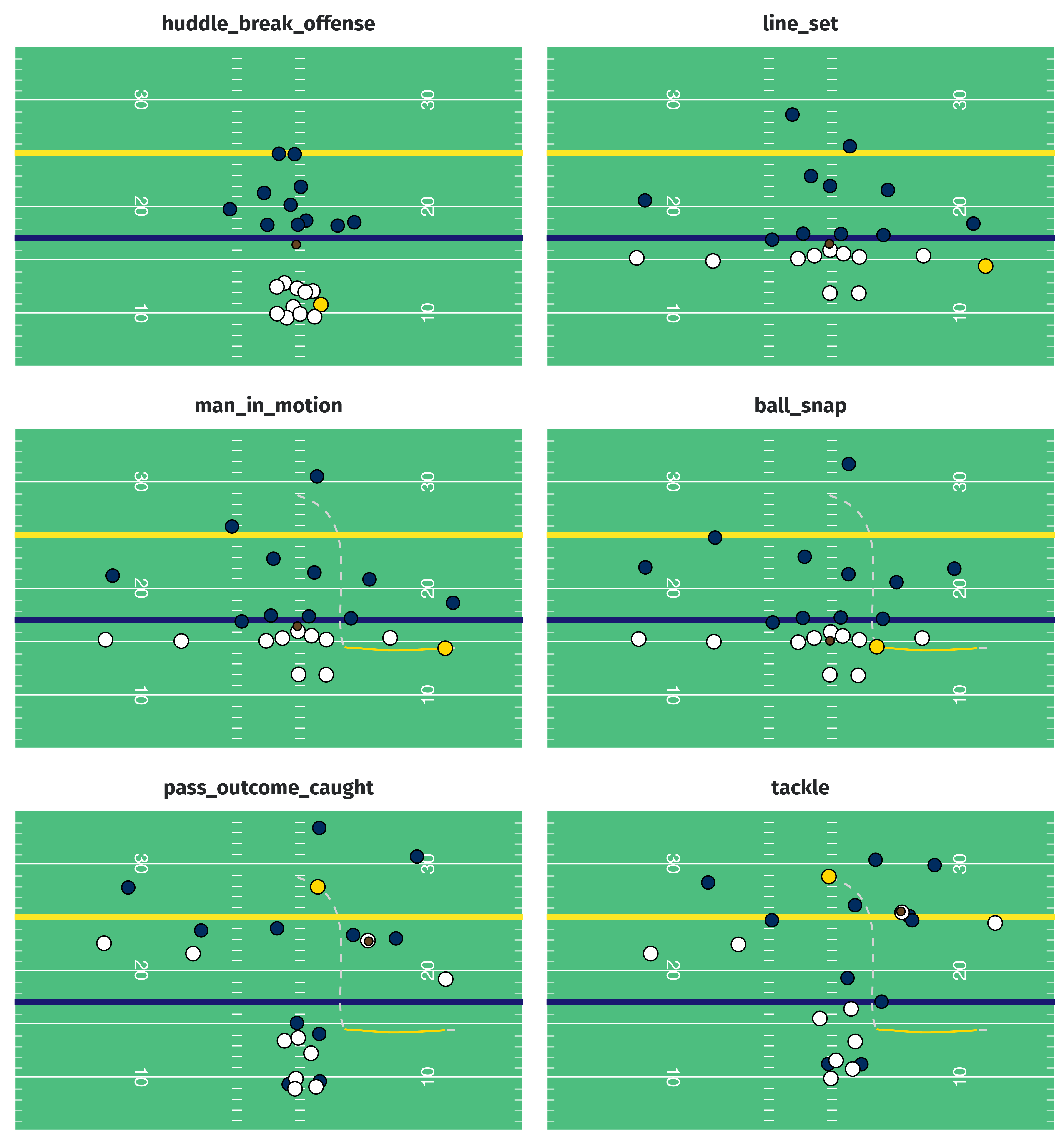} 

}

\caption{Snapshots of different events (obtained from tracking data) for a play during the Kansas City Chiefs (offense, in white) versus Indianapolis Colts (defense, in blue) game on September 25, 2022.}\label{fig:fig_field_snapshots}
\end{figure}

To provide more football context, consider the following play during the
week 3 game of the 2022 NFL regular season between the Kansas City
Chiefs and the Indianapolis Colts. Figure \ref{fig:fig_field_snapshots}
displays the locations (obtained from tracking data) of every Kansas
City (on offense, in white) and Indianapolis (on defense, in blue)
player on the field at difference events during this play. In general,
before any play begins, an offense may huddle up to strategize the play.
Once the offense breaks the huddle (\texttt{huddle\_break\_offense}), it
must officially be set at the line of scrimmage (\texttt{line\_set}).
After the line set, the offense may put a player in motion
(\texttt{man\_in\_motion}). The play then begins with the center (who
lines up in the middle of the offensive formation) snapping the football
to the quarterback (\texttt{ball\_snap}). In Figure
\ref{fig:fig_field_snapshots}, the motion player is highlighted in gold,
as he moves laterally toward the inside of the formation from an initial
outside alignment. Note that the motion player's path throughout the
play since line set is depicted with the dashed gray line, and the solid
gold line along this path represents the trajectory portion between the
start of motion and the ball snap. Once the ball is snapped, the
offensive team proceeds to execute the play. In this example, we have a
passing play, in which the quarterback throws the ball down the field to
a receiver. Here, the receiver manages to catch
(\texttt{pass\_outcome\_caught}) and advance the ball, before being
tackled by an opposing defender (\texttt{tackle}), and the play comes to
an end.

In this work, we focus on a specific aspect of pre-snap motion: the
timing between the moment a receiver starts motioning and the ball snap
event. Our main quantity of interest is the variability in snap timing,
which means across different plays, the offense does not snap the ball
at the same time after a receiver goes in motion. This can be considered
a quarterback (QB) skill, since the QB is responsible for controlling
the cadence and ensuring the offense operates smoothly in every play. As
such, it is important for the QB to synchronize the snap with motion
from their teammates, in order to keep the defense off balance and
offense in control. We posit that if the snap timing is consistent or
predictable, defenders can anticipate the snap count and time their
actions to disrupt the play. Thus, having variable snap timing may be
valuable to the offense, as it makes it harder for the opposing defense
to anticipate the snap and time their moves effectively.

Using player tracking data, we propose a Bayesian multilevel model with
heterogeneous variances to assess a quarterback's ability to manage snap
timing in the presence of pre-snap motion. In doing so, we assume a
Gamma distribution for the play-level snap timing and explicitly model
its shape parameter, allowing for player differences in the snap timing
variability. We also control for various personnel and contextual
covariates when modeling the mean snap timing, as well as random effects
for the players and teams involved. We highlight that this approach is
modular, demonstrates careful distributional consideration for the
response, and provides proper uncertainty quantification for the
estimates of all model parameters. Ultimately, our framework culminates
in a measure of variability in QB snap timing, which is demonstrated to
be predictive of the rate of havoc created by the opposing defense. We
believe our contribution provides a novel description of snap timing
variability and reveals insights into an understudied area of American
football.

The structure for the remainder of our paper is as follows. In Section
\ref{sec:data}, we offer an overview of the player tracking data in
American football. Next, we describe in detail our multilevel modeling
methodology in Section \ref{sec:methods}. We then discuss our modeling
results and insights in Section \ref{sec:results}, before closing with
our concluding remarks in Section \ref{sec:discussion}.

\section{Data}\label{sec:data}

In the analysis that follows, we leverage player tracking data from the
NFL Big Data Bowl 2025 competition (\citeproc{ref-lopez2024nfl}{Lopez et
al. 2024}), which span the first nine week of the 2022 NFL regular
season. The data are collected at a rate of 10 Hz (i.e., 10 measurements
per second) and provide two-dimensional location for every player on the
field throughout a play, along with movement attributes such as speed,
acceleration, orientation, and angle of motion. It is worth noting that
unlike previous editions of the Big Data Bowl where only a subset of
frames are included for each play, we now have access to player tracking
information for all frames within a play from pre-snap to post-snap.

The tracking data also record event tags for specific frames within each
play. Table \ref{tab:tracking} provides an example of the player
tracking data for the previously mentioned play in Section
\ref{sec:introduction}, which occurs during the game between the Kansas
City Chiefs and Indianapolis Colts in week 3 of the 2022 NFL season. For
more context, this play results in a 9-yard completed pass by Chiefs
quarterback Patrick Mahomes to wide receiver Marquez Valdes-Scantling,
with tight end Travis Kelce being the player going in motion before the
snap. In this example, the event annotations can be categorized into
three different types: pre-snap (\texttt{huddle\_break\_offense},
\texttt{line\_set}, \texttt{man\_in\_motion}), at the snap
(\texttt{ball\_snap}), and post-snap (\texttt{pass\_forward},
\texttt{pass\_outcome\_caught}, \texttt{tackle}).

Besides the tracking data, we also have access to charting data from the
NFL and Pro Football Focus. This includes statistics at the player-play
level, such as indicators about motion (e.g., whether the player goes in
motion at ball snap, shifts since the lineset, etc.), as well as both
the offense (e.g., whether the receiver is running a route, type of
route ran, etc.) and defense (e.g., whether the defender records a
quarterback hit, tackle, interception, fumble, etc.). Later on, we use
these player-play indicators to identify motion in Section
\ref{sec:definition}, and compute a summary of defensive havoc rate in
Section \ref{sec:havoc}.

\begin{table}[tbp]
\caption{Example of tracking data for a play during the Kansas City Chiefs versus Indianapolis Colts game on September 25, 2022. The data shown here are for Chiefs tight end Travis Kelce, who is the receiver in motion in this play. The columns (from left to right) are frame identifier, x-coordinate, y-coordinate, speed, acceleration, distance traveled from the frame before, orientation, angle of motion, and event annotation for each frame.
\label{tab:tracking}}
\centering
\begin{tabular}{ccccccccl}
\hline
frameId & x & y & s & a & dis & o & dir & event \\ 
\hline
1 & 20.78 & 21.18 & 2.81 & 1.46 & 0.27 & 149.05 & 161.56 & \texttt{huddle\_break\_offense} \\
\vdots & \vdots & \vdots & \vdots & \vdots & \vdots & \vdots & \vdots & \vdots \\
59 & 24.39 & 7.25 & 0.14 & 0.42 & 0.01 & 103.95 & 221.38 & \texttt{line\_set} \\
\vdots & \vdots & \vdots & \vdots & \vdots & \vdots & \vdots & \vdots & \vdots \\
81 & 24.38 & 8.13 & 1.58 & 7.33 & 0.13 & 110.90 & 358.39 & \texttt{man\_in\_motion} \\ 
\vdots & \vdots & \vdots & \vdots & \vdots & \vdots & \vdots & \vdots & \vdots \\
117 & 24.52 & 18.67 & 1.00 & 1.62 & 0.10 & 104.73 & 36.28 & \texttt{ball\_snap} \\ 
\vdots & \vdots & \vdots & \vdots & \vdots & \vdots & \vdots & \vdots & \vdots \\
138 & 34.05 & 19.18 & 6.22 & 2.34 & 0.63 & 92.17 & 83.35 & \texttt{pass\_forward} \\ 
\vdots & \vdots & \vdots & \vdots & \vdots & \vdots & \vdots & \vdots & \vdots \\
147 & 37.83 & 21.50 & 4.25 & 3.67 & 0.43 & 289.61 & 32.34 & \texttt{pass\_outcome\_caught} \\ 
\vdots & \vdots & \vdots & \vdots & \vdots & \vdots & \vdots & \vdots & \vdots \\
157 & 38.80 & 23.70 & 0.79 & 2.56 & 0.09 & 208.67 & 43.25 & \texttt{tackle} \\ 
\hline
\end{tabular}
\end{table}

\section{Methods}\label{sec:methods}

\subsection{Defining snap timing}\label{sec:definition}

In our analysis, we focus on passing plays with receivers in motion at
the ball snap and running a route, resulting in a total of 2,198 plays.
For play \(i = 1, \dots, n\), let \(t_i^{\text{motion}}\) denote the
moment a receiver starts going in motion, and \(t_i^{\text{snap}}\)
denote the moment of ball snap. We define the snap timing for play \(i\)
as the number of frames elapsed between \(t_i^{\text{motion}}\) and
\(t_i^{\text{snap}}\). That is,
\[\delta_i = t_i^{\text{snap}}-t_i^{\text{motion}}.\] Here, it is
straightforward to obtain \(t_i^{\text{snap}}\) using the frame label in
the tracking data. Conversely, identifying \(t_i^{\text{motion}}\)
requires more effort, due to inconsistency between player-level motion
indicators and annotated tracking events, as acknowledged by the
competition organizers\footnote{See
  \href{https://www.kaggle.com/competitions/nfl-big-data-bowl-2025/discussion/543709}{\texttt{https://www.kaggle.com/competitions/nfl-big-data-bowl-2025/discussion/543709}}.}.

We use the following procedure to determine \(t_i^{\text{motion}}\) for
each play.

\begin{itemize}
\item
  For plays where the player in motion at the ball snap is also the only
  player in motion since line set, we identify \(t_i^{\text{motion}}\)
  as the frame of the \texttt{man\_in\_motion} event provided in the
  tracking data.
\item
  For the remaining plays (specifically, plays where the motion player
  at the ball snap is not the only player in motion since line set, and
  plays with players charted as in motion at the snap but without a
  \texttt{man\_in\_motion} event):

  \begin{itemize}
  \item
    Using plays with \(t_i^{\text{motion}}\) identified from earlier, we
    first observe the distribution for the ratio between the motion
    player's speed at \texttt{man\_in\_motion} and their maximum speed
    between line set and ball snap (see Figure
    \ref{fig:fig_ratio_distribution} in the Supplementary Materials).
  \item
    We then choose the average ratio value of 0.45 as the threshold for
    determining the start of motion for the remaining plays.
    Specifically, we assign \(t_i^{\text{motion}}\) as the first frame
    between line set and ball snap where the player reaches at least
    45\% of their top speed.
  \end{itemize}
\end{itemize}

\begin{figure}[H]

{\centering \includegraphics{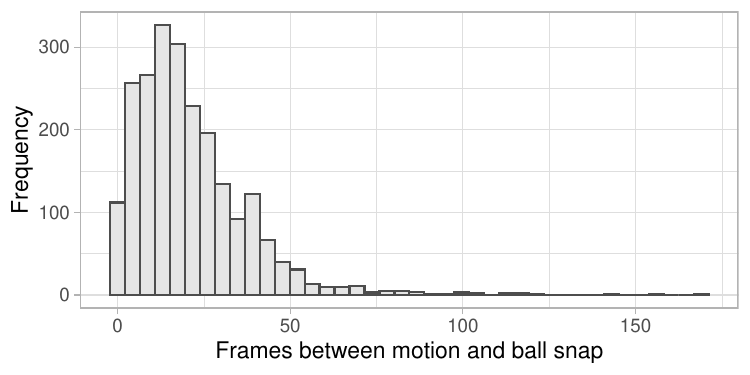} 

}

\caption{Distribution of the snap timing across plays with receiver going in motion at the snap and running a route.}\label{fig:fig_timing_distribution}
\end{figure}

For context, Figure \ref{fig:fig_timing_distribution} displays the snap
timing distribution for plays with receivers in motion at the ball snap
and running a route. We observe a right skewed distribution, as about
95\% of the snap timing values are within 50 frames (or 5 seconds), with
a median of 17 frames (i.e., 1.7 seconds between the start of motion and
ball snap). It appears that the snap timing is relatively short across
the considered motion plays.

\subsection{Multilevel model for snap timing}\label{sec:model}

The observed timing between motion and snap on a play is likely
attributable to numerous variables, from the play-level context to the
players and teams involved. Apart from the quarterback and receiver on
offense, there is variability in the opposing team on defense. Thus, it
is necessary to decompose the variability in snap timing into factors
related to the individual player and team on any given play.

To this end, we fit the following multilevel model: \[
\begin{aligned}
\delta_i&\sim\textsf{Gamma}(\mu_i,\alpha_i)\\
\\
\log\mu_i&=\gamma_0+\boldsymbol{\beta X_i}+b_{q[i]}+b_{m[i]}+b_{d[i]}\\
b_q&\sim\textsf{Normal}(0,\sigma^2_q)\\
b_m&\sim\textsf{Normal}(0,\sigma^2_m)\\
b_d&\sim\textsf{Normal}(0,\sigma^2_d)\\
\\
\log\alpha_i&=\psi_0+u_{q[i]}\\
u_q&\sim\textsf{Normal}(0,\tau^2_q)\\
\end{aligned}
\]

In detail, we model the response---the snap timing for play \(i\)
(\(\delta_i>0\))---using a Gamma distribution, parameterized in terms of
its mean \(\mu\) and the shape parameter \(\alpha > 0\). This is
well-suited for modeling a positive, continuous (``interarrival time'')
variable like our case. With this specification, we can fit separate
models for both parameters and see how the overall distribution shifts
based on different explanatory variables.

Of primary interest, we model the shape parameter \(\alpha\) (which is
proportional to the Gamma distribution's variance) with QB random
effects. This enables us to estimate the differences in the snap timing
variability among NFL quarterbacks. From an offensive standpoint, we
hypothesize that a higher timing variability is beneficial because it
prevents defenses from predicting when the snap will occur. This may
allow the offense to control the game flow and exploit defensive
vulnerabilities.

We also model the mean \(\mu\) by including random intercepts for three
groups: quarterback \(q\), motion player \(m\), and defensive team
\(d\). Moreover, we account for covariate information about play \(i\)
through \(\boldsymbol{X_i}\), and estimate the coefficients
\(\boldsymbol\beta\) as fixed effects. In particular, we adjust for
contextual information (down, play clock, and timeouts remaining) since
they can dictate the urgency and type of play selected, which in turn
can affect snap timing.

For instance, on earlier downs, offenses typically have more flexibility
and may take additional time before letting a player go in motion and
snapping the ball. The availability of timeouts can also relate to how
teams manage snap timing, as it impacts the offense's strategies to
either conserve or consume time based on the game situation. Further, to
account for any positional effects, we control for the position of the
motion player (running back, tight end, and wide receiver), with running
back being the reference level. The number of players in motion since
line set, which can be summarized using the player-play data mentioned
in Section \ref{sec:data}, can also have an impact the timing between
motion and snap.

Additionally, the type of motion by receiver is a vital feature for
modeling snap timing, since it relates to play design. For example, in a
play-action pass, a team can fake a jet sweep to set up the quarterback
for a pass. Or, a team may also employ the jet motion to bring a
receiver from one side of the formation to the other to create a
favorable matchup. This information, unlike the other variables, is not
provided in the data, prompting us to consider a clustering algorithm to
identify the different types of motion in Section \ref{sec:clustering}.

We fit the multilevel model for snap timing in a Bayesian framework via
the \texttt{brms} package in \texttt{R}
(\citeproc{ref-burkner2017brms}{Bürkner 2017},
\citeproc{ref-burkner2018advanced}{2018},
\citeproc{ref-burkner2021bayesian}{2021}), which provides an interface
for Bayesian modeling with \texttt{Stan}
(\citeproc{ref-carpenter2017stan}{Carpenter et al. 2017}). We use the
following weakly informative prior distributions for the standard
deviation parameters in our model: \[
\begin{aligned}
\sigma_q &\sim \textsf{half-t}_3\\
\sigma_m &\sim \textsf{half-t}_3\\
\sigma_d &\sim \textsf{half-t}_3\\
\tau_q &\sim \textsf{half-t}_3\\
\end{aligned}
\] where \(\textsf{half-t}_3\) represents a half-\(t\) distribution
(i.e., the absolute value of a Student-\(t\) distribution centered at
zero) with 3 degrees of freedom (see Gelman
(\citeproc{ref-gelman2006prior}{2006}) for more details).

Our Bayesian approach naturally provides uncertainty quantification for
the model parameters via their posterior distributions, which are
estimated using MCMC sampling. For model fitting, we use 4 parallel
chains, each having 10,000 iterations and 5,000 warmup draws. We assess
the convergence of the MCMC algorithm with trace plots and the
\(\widehat{R}\) statistic (\citeproc{ref-gelman1992inference}{Gelman and
Rubin 1992}; \citeproc{ref-brooks1998general}{Brooks and Gelman 1998}).
We find good evidence of convergence after visual inspection of the
trace plots, as also supported by all \(\widehat{R}\) values being very
close to 1. We also compute the effective sample size for each parameter
(\citeproc{ref-gelman2013bayesian}{Gelman et al. 2013}) and observe no
issues with model efficiency.

\subsection{Motion type clustering}\label{sec:clustering}

As alluded to in Section \ref{sec:model}, we aim to perform a cluster
analysis to provide unsupervised labels for the type of motion by NFL
receivers. This ultimately allows us to include the identified motion
clusters as a covariate in our multilevel model for snap timing. To do
so, we first derive features characterizing the locations and
trajectories of motion players at various points during a play, so that
a clustering of these variables will output a meaningful interpretation
of the motion type.

Our clustering input contains tracking data features describing the
receiver locations at different on-field events in relation to where the
center lines up before the snap. To some extent, these space-time
attributes should reasonably summarize the magnitude and direction of
the motion player's trajectory, capturing their forward and lateral
movements throughout a play. Note that our motion type detection
strategy is similar to the approach in Dutta, Yurko, and Ventura
(\citeproc{ref-dutta2020unsupervised}{2020}) for identifying defensive
coverage types. That is, we extract features based on the
high-resolution tracking data before performing model-based clustering
to provide event annotations. In general, this can be applied to various
unsupervised labeling problems in sports, given the availability of
tracking data.

\begin{figure}[ptb]

{\centering \includegraphics[width=0.7\linewidth]{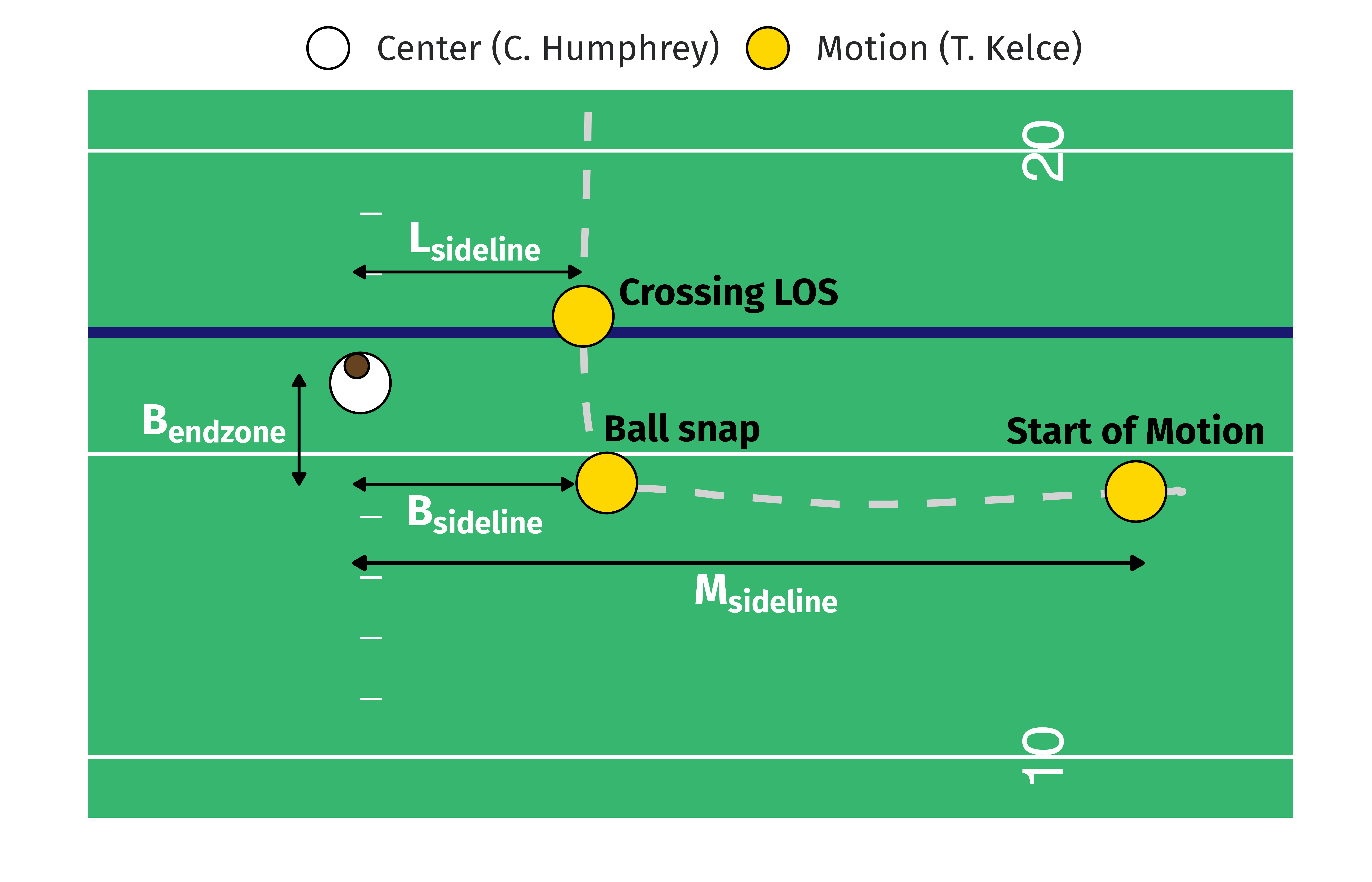} 

}

\caption{A diagram showing the features used for clustering the type of motion. The motion player is Travis Kelce, color-coded gold, and the center is Creed Humphrey, color-coded white. The derived features are the displacement between the center and motion player at ball snap in both sideline $(\textsf{B}_{\textsf{sideline}})$ and endzone $(\textsf{B}_{\textsf{endzone}})$ directions, as well as the displacement with respect to the sideline between the center and motion player at the start of motion $(\textsf{M}_{\textsf{sideline}})$ and when the motion player crosses the line of scrimmage $(\textsf{L}_{\textsf{sideline}})$.}\label{fig:fig_clustering_features}
\end{figure}

For illustration, Figure \ref{fig:fig_clustering_features} shows the
derived features for the example play with Travis Kelce going in motion
as described in Table \ref{tab:tracking}. In particular, we compute the
following attributes:

\begin{itemize}
\item
  Displacement between the center and motion player at ball snap with
  respect to the sideline \((\textsf{B}_{\textsf{sideline}})\)
\item
  Displacement between the center and motion player at ball snap with
  respect to the target endzone \((\textsf{B}_{\textsf{endzone}})\)
\item
  Displacement between the center and motion player at the start of
  motion with respect to the sideline
  \((\textsf{M}_{\textsf{sideline}})\)
\item
  Displacement between the center and motion player at the moment of
  crossing the line of scrimmage with respect to the sideline
  \((\textsf{L}_{\textsf{sideline}})\)
\end{itemize}

Note that for plays where the motion player never crosses the line of
scrimmage, we instead use their location at either the quarterback event
(e.g., forward pass, sack, etc.) or 3 seconds after the snap, depending
on which happens sooner, in place of \(\textsf{L}_{\textsf{sideline}}\).
The aforementioned 3-second threshold is chosen using the observed
distribution of the time elapsed from snap to crossing the line of
scrimmage for plays where motion players do cross the line of scrimmage,
as it captures a sufficiently large fraction of the values (see Figure
\ref{fig:fig_snap_cross_los_distribution} in the Supplementary
Materials).

To obtain unsupervised labels for the pre-snap motion types, we consider
a Gaussian mixture model (\citeproc{ref-banfield1993model}{Banfield and
Raftery 1993}). As a model-based clustering method, Gaussian mixture
model provides a density-based statistical approach for identifying
groups of observations, as opposed to heuristic-based algorithms that
detect the clusters directly based on the data. The algorithm fits a
mixture of Gaussian densities to the data, where each density is
representative of an individual component (or ``cluster''). Moreover,
the method returns soft cluster assignments, which enables uncertainty
quantification when assigning cluster membership probabilities to
observations. For a complete survey of model-based clustering, we refer
the reader to Gormley, Murphy, and Raftery
(\citeproc{ref-gormley2023model}{2023}).

We fit a Gaussian mixture model to detect the types of motion by NFL
receivers using the \texttt{mclust} package in \texttt{R}
(\citeproc{ref-scrucca2023model}{Scrucca et al. 2023};
\citeproc{ref-r2024language}{R Core Team 2024}), which performs maximum
likelihood estimation via the EM algorithm
(\citeproc{ref-dempster1977maximum}{Dempster, Laird, and Rubin 1977}).
We use the Bayesian information criterion
(\citeproc{ref-schwarz1978estimating}{Schwarz 1978}) to select the
optimal model and number of Gaussian clusters \(G^*\)
(\citeproc{ref-raftery2006variable}{Raftery and Dean 2006}). After
fitting, a mixture of \(G^*=6\) ellipsoidal multivariate Gaussian
distributions with varying volume, shape, and orientation (VVV) is
deemed the best model. We further validate our results through visual
inspection of motion trajectories of players from the 6 identified
clusters (see Section \ref{sec:clustering-supp} of the Supplementary
Materials).

We emphasize that our approach for providing annotations for the motion
types is only a starting point and should serve as a foundation for
future work. We note that future improvements regarding feature
engineering and unsupervised learning algorithm may produce more
reasonable clustering results, which will be discussed in Section
\ref{sec:discussion}. Ultimately, we condition on these cluster labels
as fixed effects in modeling the mean parameter \(\mu\), so that our
estimates of QB snap timing variability \(u_q\) at the shape level are
not confounded by the different types of motion.

\section{Results}\label{sec:results}

\subsection{Model summary}\label{model-summary}

We now examine the estimates of the fixed effects coefficients and
variance parameters in our model. Table \ref{tab:fixedeff} summarizes
the results for the fixed effects terms of the mean snap timing model in
our multilevel framework. We notice several interesting results
regarding the play-level contextual and personnel covariates.

Relative to first down plays, the timing between the start of motion by
a receiver and the ball snap appears to take longer on second and third
downs. This could be due to more complex offensive play calls on later
downs, or the defense's tendency to disguise their coverage more on
crucial downs, requiring the quarterback to make last-second reads and
adjustments. We also see that on average, the snap timing tends to be
longer when there are no timeouts left for the offense. The lacks of
timeouts can affect the offense to be more deliberate in execution,
hence taking more time between motion and snap than when there are one
or more timeouts available. In addition, we observe differences among
the receiver positions, as plays with running backs going in motion have
shorter snap timing than for tight ends and wide receivers. This is
perhaps expected based on where each position usually lines up in a
formation before the snap. Specifically, running backs are typically in
the middle, close to the center and quarterback, whereas wide receivers
often line up in an outside wide alignment.

\begin{table}[tbp]
\caption{Posterior estimates for the fixed effects coefficients (at the mean level) for the multilevel model for snap timing. Note that the reference down level is first down, denoted down:1; the reference number of timeouts remaining level is 0, denoted timeouts:0; the reference motion player's position level is running back, denoted position:RB; and the reference motion type level (obtained via clustering) is the first motion cluster label, denoted motion:1. \label{tab:fixedeff}}
\centering
\begin{tabular}{lrrrr}
\hline
\multicolumn{1}{c}{} & \multicolumn{2}{c}{Posterior summary} & \multicolumn{2}{c}{95\% credible interval} \\
& Mean & SD & Lower & Upper \\ 
\hline
Intercept & $2.409$ & $0.131$ & $2.157$ & $2.670$ \\ 
$I_{\text{\{down:2\}}}$ & $0.068$ & $0.034$ & $0.000$ & $0.134$ \\
$I_{\text{\{down:3\}}}$ & $0.192$ & $0.041$ & $0.113$ & $0.272$ \\
$I_{\text{\{down:4\}}}$ & $0.150$ & $0.113$ & $-0.068$ & $0.375$ \\
Play clock at motion & $0.029$ & $0.003$ & $0.022$ & $0.035$ \\ 
$I_{\text{\{timeouts:1\}}}$ & $-0.278$ & $0.131$ & $-0.534$ & $-0.024$ \\
$I_{\text{\{timeouts:2\}}}$ & $-0.143$ & $0.112$ & $-0.367$ & $0.073$ \\
$I_{\text{\{timeouts:3\}}}$ & $-0.151$ & $0.108$ & $-0.369$ & $0.054$ \\
$I_{\text{\{position:TE\}}}$ & $0.531$ & $0.065$ & $0.405$ & $0.659$ \\
$I_{\text{\{position:WR\}}}$ & $0.320$ & $0.054$ & $0.214$ & $0.426$ \\
Motion players since line set & $-0.021$ & $0.032$ & $-0.084$ & $0.042$ \\
$I_{\text{\{motion:2\}}}$ & $0.039$ & $0.053$ & $-0.065$ & $0.144$ \\
$I_{\text{\{motion:3\}}}$ & $0.560$ & $0.050$ & $0.463$ & $0.659$ \\
$I_{\text{\{motion:4\}}}$ & $-0.029$ & $0.046$ & $-0.118$ & $0.061$ \\
$I_{\text{\{motion:5\}}}$ & $-0.192$ & $0.072$ & $-0.335$ & $-0.049$ \\
$I_{\text{\{motion:6\}}}$ & $0.154$ & $0.064$ & $0.029$ & $0.281$ \\
\hline
\end{tabular}
\end{table}

\begin{table}[tbp]
\caption{Posterior estimates for the standard deviation of the random effect terms in the multilevel model for snap timing. \label{tab:randeff}}
\centering
\begin{tabular}{lrrrr}
\hline
\multicolumn{1}{c}{} & \multicolumn{2}{c}{Posterior summary} & \multicolumn{2}{c}{95\% credible interval} \\
Parameter & Mean & SD & Lower & Upper \\ 
\hline
$\sigma_q$ & $0.093$ & $0.031$ & $0.027$ & $0.155$ \\
$\sigma_m$ & $0.151$ & $0.031$ & $0.088$ & $0.210$ \\
$\sigma_d$ & $0.029$ & $0.020$ & $0.001$ & $0.074$ \\
$\tau_q$ & $0.297$ & $0.052$ & $0.205$ & $0.409$ \\
\hline
\end{tabular}
\end{table}

Next, Table \ref{tab:randeff} presents posterior estimates for the
standard deviation of the random effect terms in our multilevel model.
Among the estimates when modeling the mean snap timing \(\mu\), we
observe the highest posterior mean for the standard deviation of motion
players (\(\hat \sigma_m = 0.151\)), followed by quarterbacks
(\(\hat \sigma_q = 0.093\)) and defenses (\(\hat \sigma_d = 0.029\)).
Note that the standard deviation estimates for the two offensive groups
are farther away from zero than the defensive team's, suggesting some
level of variability between players within each position group. For
both motion players and quarterbacks, we provide analyses of their
random effects \(b_q\) and \(b_m\) in Section \ref{sec:model-supp} of
the Supplementary Materials, with Figures
\ref{fig:fig_qb_mean_posterior_distributions} and
\ref{fig:fig_receiver_posterior_distributions} showing their respective
posterior distributions.

More importantly, across all considered random effects, the largest
source variation is captured by \(\hat \tau_q = 0.297\). Thus, the
highest standard deviation estimate is between quarterbacks when
modeling the snap timing shape \(\alpha\), compared to the rest of the
estimates in the mean parameter model. This leads us to focus on the QB
shape random effects \(u_q\) in the following analysis.

\subsection{Quarterback leaderboard}\label{sec:leaderboard}

\begin{figure}[ptb]

{\centering \includegraphics[width=0.75\linewidth]{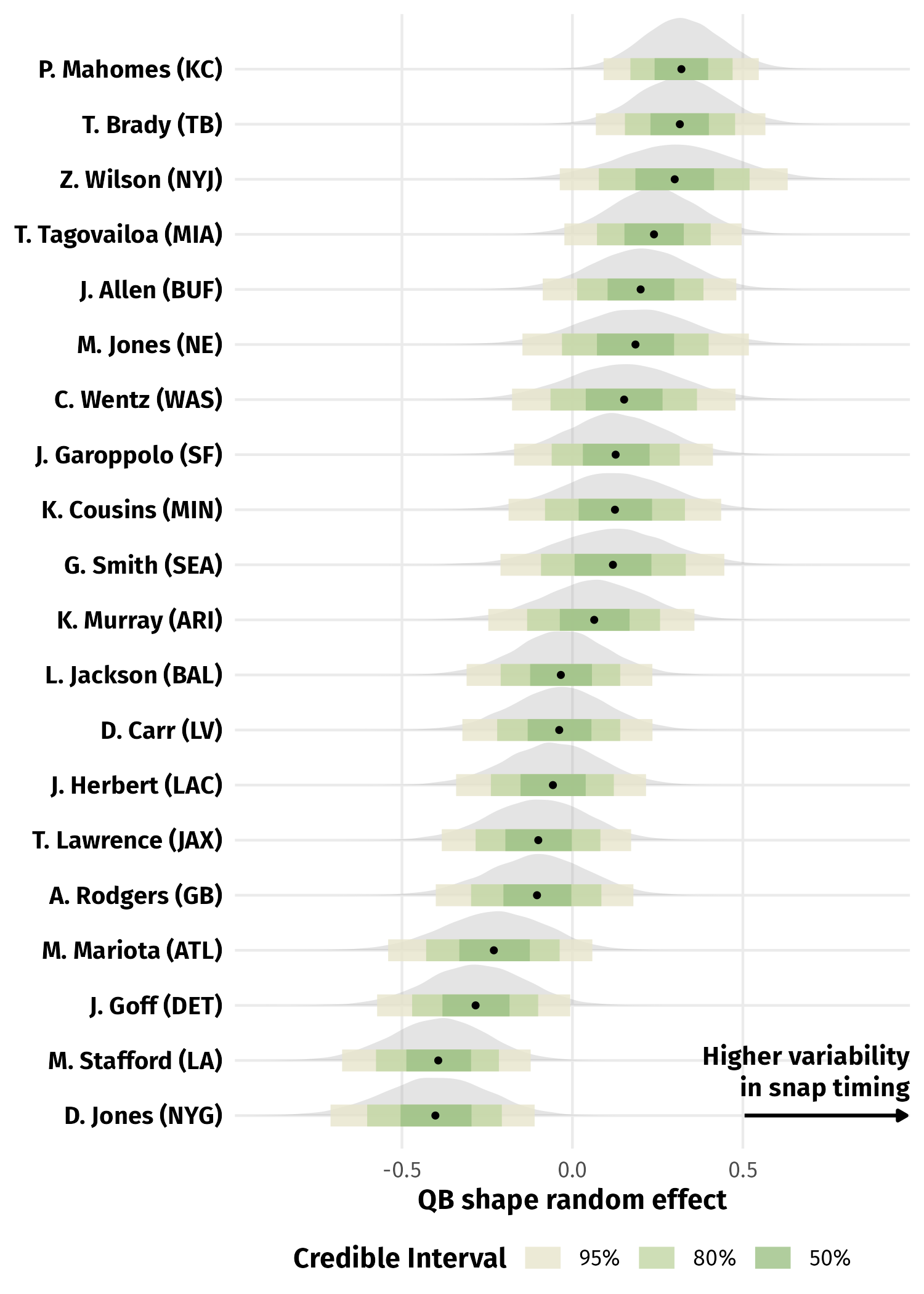} 

}

\caption{Posterior distributions of the shape random effect for NFL quarterbacks with at least 50 pass attempts on plays with receivers in motion at snap and running a route. For each player, the posterior mean point estimate and corresponding credible intervals are depicted. Here, a higher posterior mean demonstrates greater variability in snap timing.}\label{fig:fig_posterior_distributions}
\end{figure}

Figure \ref{fig:fig_posterior_distributions} displays the posterior
distributions for the shape random effect \(u_q\) for quarterbacks with
at least 50 pass attempts across the considered motion plays. Here, the
player ordering is based on the posterior mean estimate, which captures
a quarterback's ability to maintain variable snap timing. More
specifically, a higher posterior mean values corresponds to greater
variability in timing between the start of motion and snap.

At the top of our leaderboard, Patrick Mahomes stands out as the highest
ranked QB according to our measure, adding the ability to time snaps as
another element to his success. Other high-caliber quarterbacks such as
Tom Brady and Josh Allen are also among the leaders in our rankings.
Note that there is considerable uncertainty in our estimates, as
demonstrated by the rather wide credible intervals. We also observe that
most of the posterior distributions are not entirely above or below
zero. This is unsurprising given our limited sample of motion plays.

Still, there are notable differences in the credible intervals among
this subset of quarterbacks. For instance, the 95\% credible interval
for Daniel Jones (who ranks last in our leaderboard) does not overlap
with the top five quarterbacks' 95\% intervals. This indicates that our
posterior mean estimates provide discriminative power for distinguishing
between the players within this limited subset of data.

\subsection{Relationship between variability in snap timing and havoc
rate}\label{sec:havoc}

\begin{figure}[ptb]

{\centering \includegraphics[width=1\linewidth]{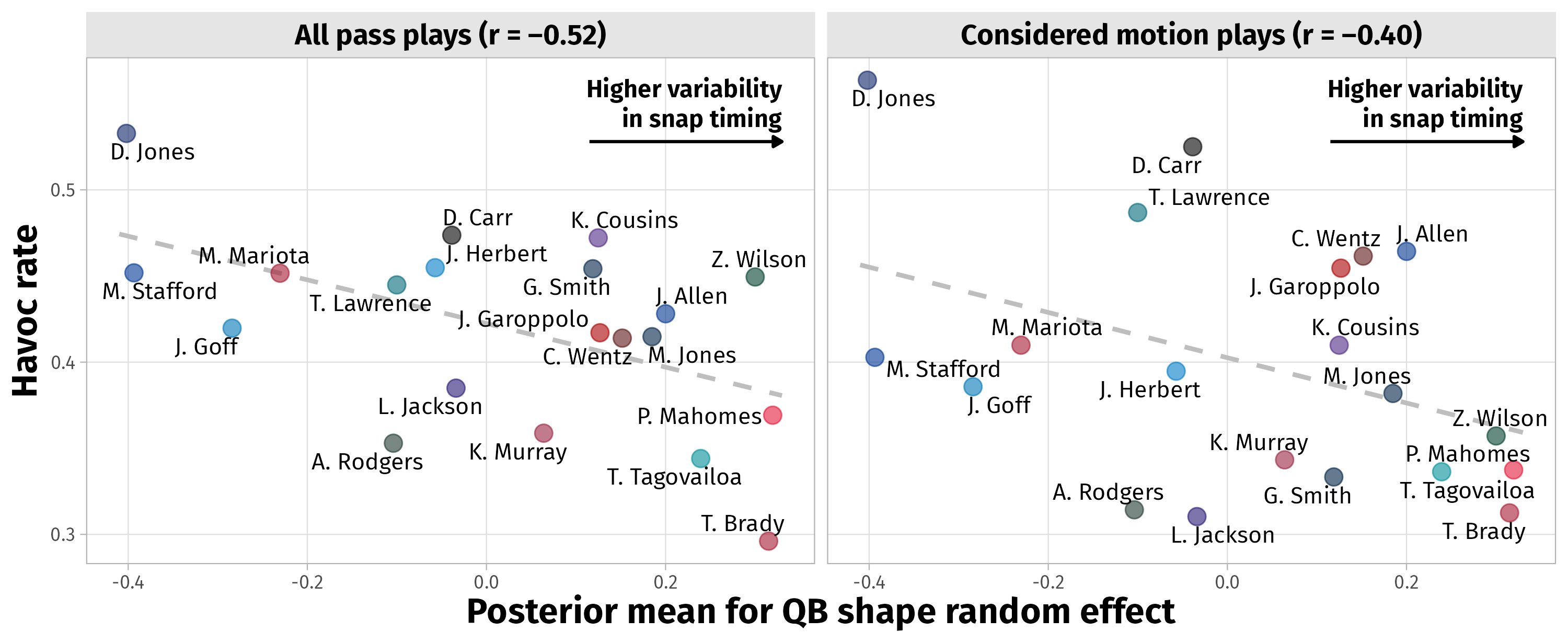} 

}

\caption{Relationship between the posterior mean of the QB shape random effect and the rate of havoc generated by the opposing defense across for all passing plays in the first nine weeks of the 2022 NFL season (left) and only the considered motion plays (right). Results shown here are for NFL quarterbacks with at least 50 pass attempts on plays with receivers in motion at snap and running a route.}\label{fig:fig_corr_havoc}
\end{figure}

We now examine the relationship between the posterior mean for the QB
shape random effect \(u_q\) and a measure of play-level havoc rate.
Here, a havoc event is defined as whether any of the following defensive
outcomes is generated on a play: pass breakup, forced fumble, tackle for
loss, interception, sack, and pressure---each of which can be summarized
using the player-play information described in Section \ref{sec:data}.
Figure \ref{fig:fig_corr_havoc} displays scatterplots of our posterior
mean estimates for the same subset of QBs as before and (left) the havoc
rate across all passing plays over the first nine weeks of the 2022 NFL
season and (right) the havoc rate for only the considered motion plays
in our analysis.

We observe that lower snap timing variability corresponds to higher rate
of facing havoc events created by the opposing defense. This makes
intuitive sense, as when there is little variability in timing, the
offense is likely to be predictable and experience unfavorable play
outcomes. In contrast, by varying the duration between motion and snap,
offenses can create uncertainty, forcing defenders to play more
cautiously which reduces their effectiveness in executing disruptive
plays. Note that although we focus on snap timing for passing plays with
motion, our estimates may also serve as proxies for QB awareness or
pocket presence, as indicated by the moderate correlation with havoc
rate across all passing plays, not just the plays considered in our
model.

As a side note, the posterior mean for the shape random effect scarcely
correlates with the rate of motion on all passing plays during the first
nine weeks of the 2022 season (\(r=0.11\); see Figure
\ref{fig:fig_corr_shape_motion} in the Supplementary Materials). Over
this period, there is also a weak correlation between the motion rate
and havoc rate across all passing plays (\(r=-0.09\); see Figure
\ref{fig:fig_corr_motion_havoc} in the Supplementary Materials). This
suggests that our measure of QB snap timing variability provides
independent information of motion tendency, while also being a more
direct indicator of encountering defensive disruptions than the rate of
motion.

\section{Discussion}\label{sec:discussion}

Variability in snap timing is an essential aspect for an offense to
dictate plays and make it difficult for defenses to anticipate and
react. Through multilevel modeling, we provide an assessment of a
quarterback's ability to maintain variable snap timing with their
teammates' motion on passing plays. We assume a Gamma distribution for
our outcome variable---the snap timing on a play---and account for
relevant fixed effects to capture mean shifts in snap timing along with
random effects for the quarterback, motion receiver, and opposing
defense. We also include QB random effects to model the shape parameter
of the Gamma distribution, enabling us to estimate the differences in
snap timing variability between NFL quarterbacks. Since our model is fit
in a Bayesian framework, it allows us to quantify uncertainty with
posterior distributions of all model parameters. Our results suggest
that higher variability in snap timing is beneficial for the passing
game, as it relates to experiencing less disruptions generated by the
defense.

Our proposed framework is not without limitations. First, to identify
the start of motion, we use a threshold-based criterion which relies on
the moment the motion player achieves a certain fraction of their top
speed before the snap. This is certainly a simple viewpoint, as one
could come up with a more robust definition by building a predictive
model to detect the starting motion frame. Second, despite the
reasonable set of inputs in our cluster analysis, there is room for
improvement in identifying the motion types. It is possible to derive
more refined contextual and tracking features, or alternatively consider
a functional clustering approach. We highlight that the current modeling
framework is modular, so that our simple algorithm in Section
\ref{sec:clustering} can be replaced by other approaches for clustering
the type of motion. For example, one could adapt a model-based curve
clustering strategy similar to what Chu et al.
(\citeproc{ref-chu2020route}{2020}) use for route identification, to the
context of motion types of receivers. Since neither of these tasks is
the main focus of our work, we leave these explorations for the future.

Additionally, our study is limited to only passing plays with receivers
in motion at the ball snap. This introduces a selection bias into our
analysis by disregarding running plays in the provided data sample.
Moreover, we exclude various quarterbacks who do not meet the cutoff for
minimum pass attempts on motion plays (e.g., Jalen Hurts and Joe
Burrow). We also recognize that while our estimates are referred to as
\emph{QB random effects}, they are certainly correlated with the team's
center and coach's play-calling. For instance, the success on offense of
the 2022 Kansas City Chiefs is not solely due to the greatness of
Patrick Mahomes. Instead, the brilliant offensive mind of head coach
Andy Reid, as well as their starting center Creed Humphrey (named Pro
Bowl and second-team All-Pro in 2022), both play a important part in the
team's snap timing execution. We recognize that these are challenging
issues, and leave these investigations for future work.

\section*{Acknowledgements}\label{acknowledgements}
\addcontentsline{toc}{section}{Acknowledgements}

We thank the organizers of the NFL Big Data Bowl 2025 for hosting the
competition and providing access to the data, as well as Sam
Schwartzstein for a conversation which motivated the topic of
quarterback snap timing variability.

\section*{Code availability}\label{code-availability}
\addcontentsline{toc}{section}{Code availability}

All code related to this paper is available at
\href{https://github.com/qntkhvn/timing}{\texttt{https://github.com/qntkhvn/timing}}.

\section*{References}\label{references}
\addcontentsline{toc}{section}{References}

\phantomsection\label{refs}

\newpage

\def\thesection{\Alph{section}}
\counterwithin{figure}{section}
\counterwithin{table}{section}

\newcounter{sectionstoskip}
\setcounter{sectionstoskip}{13}
\addtocounter{section}{\value{sectionstoskip}}

\section{Supplementary materials}\label{supplementary-materials}

\subsection{Additional figures}\label{additional-figures}

\begin{figure}[H]

{\centering \includegraphics{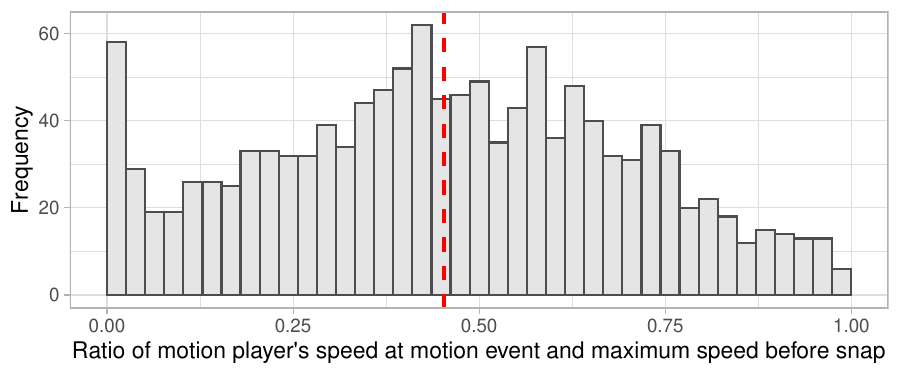} 

}

\caption{Distribution of the ratio between the motion player's speed at $\texttt{man\_in\_motion}$ event and their maximum speed between line set and ball snap (for plays where the receiver in motion at snap is the only player going in motion since line set). The red dashed line represents the average value of 0.45, which is chosen as the threshold for identifying motion for the rest of the plays.}\label{fig:fig_ratio_distribution}
\end{figure}

\begin{figure}[H]

{\centering \includegraphics{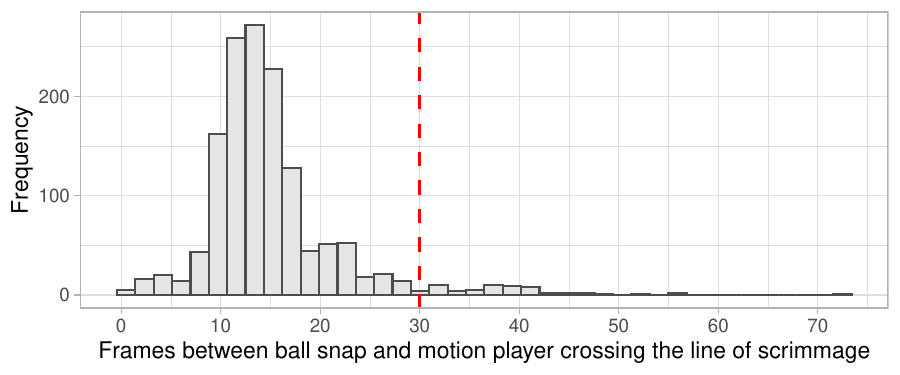} 

}

\caption{Distribution for the number of frames between the snap and motion player crossing the line of scrimmage (for plays where motion players cross the line of scrimmage). The red dashed line represents the time threshold of 30 frames (i.e., 3 seconds), which captures about 95\% of the values.}\label{fig:fig_snap_cross_los_distribution}
\end{figure}

\begin{figure}[ptb]

{\centering \includegraphics{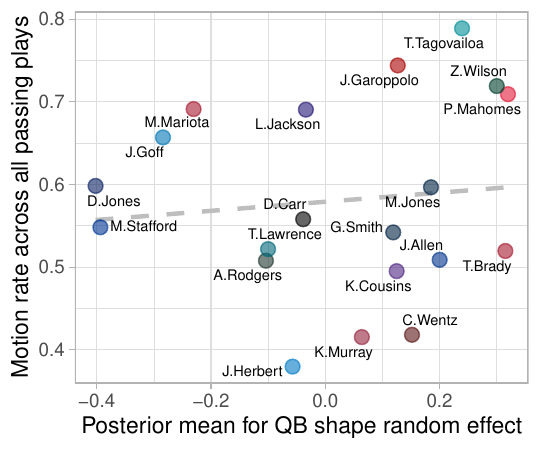} 

}

\caption{Relationship between the posterior mean of the QB shape random effect and the rate of motion by their corresponding team for all passing plays in the first nine weeks of the 2022 NFL season $(r= 0.11)$. Results shown here are for NFL quarterbacks with at least 50 pass attempts on plays with receivers in motion at snap and running a route.}\label{fig:fig_corr_shape_motion}
\end{figure}

\begin{figure}[ptb]

{\centering \includegraphics{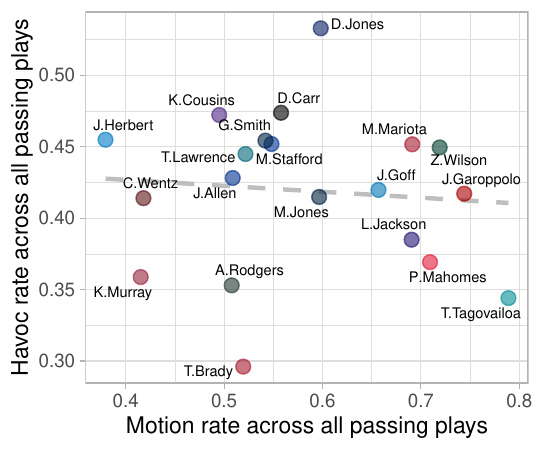} 

}

\caption{Relationship between the motion rate and havoc rate across all passing plays in the first nine weeks of the 2022 NFL season $(r= -0.09)$. Results shown here are for NFL quarterbacks with at least 50 pass attempts on plays with receivers in motion at snap and running a route.}\label{fig:fig_corr_motion_havoc}
\end{figure}

\subsection{Additional clustering results}\label{sec:clustering-supp}

\begin{figure}[H]

{\centering \includegraphics{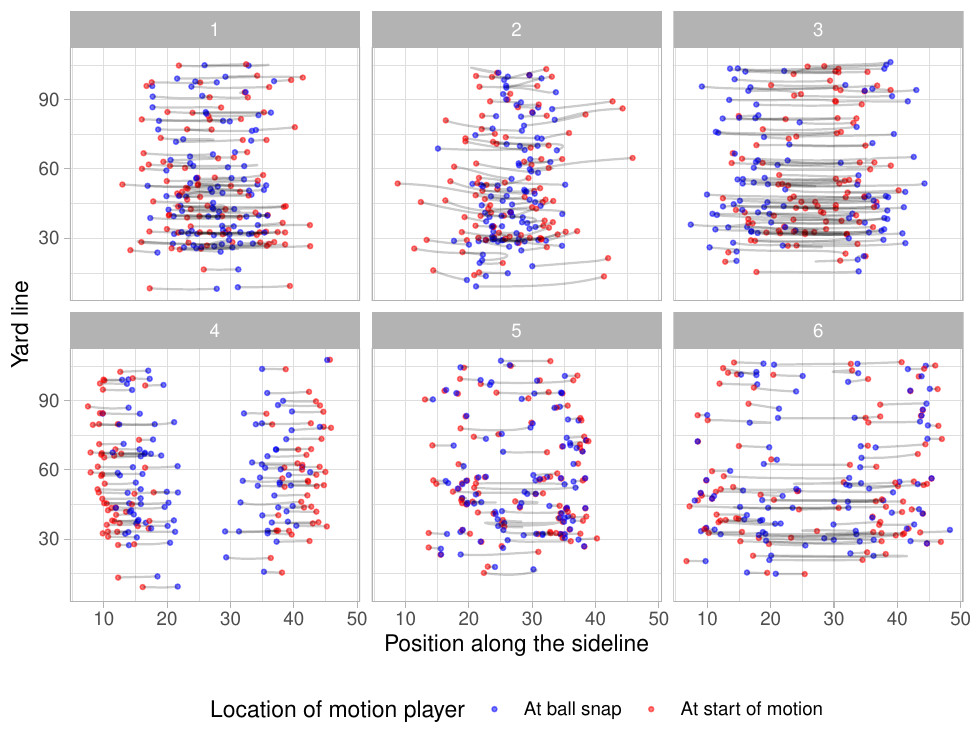} 

}

\caption{A sample of 100 player trajectories between motion and snap in each cluster label obtained from the Gaussian mixture model.}\label{fig:fig_motion_clusters}
\end{figure}

We visually inspect and analyze the results of fitting the Gaussian
mixture model for characterizing the type of receiver motion as
described in Section \ref{sec:clustering}. Figure
\ref{fig:fig_motion_clusters} displays sample paths of motion players in
each of the 6 identified clusters from the starting point of motion (in
red) to the ball snap (in blue). We observe clear structural differences
in the player paths between the motion labels, which validates our
clustering output.

We can then interpret the cluster assignments to add more football
context to the results. For instance, cluster 2 represents the orbit
motion, as the motion happens behind the quarterback with the motion
player moving laterally until the snap; cluster 3 illustrates the jet
motion, where the motion player runs laterally across the formation;
cluster 4 depicts the glide motion, with the motion player initially
lining up in an outside wide alignment before moving laterally toward
the inside; and so on.

\subsection{Additional multilevel model results}\label{sec:model-supp}

\begin{figure}[ptb]

{\centering \includegraphics[width=0.75\linewidth]{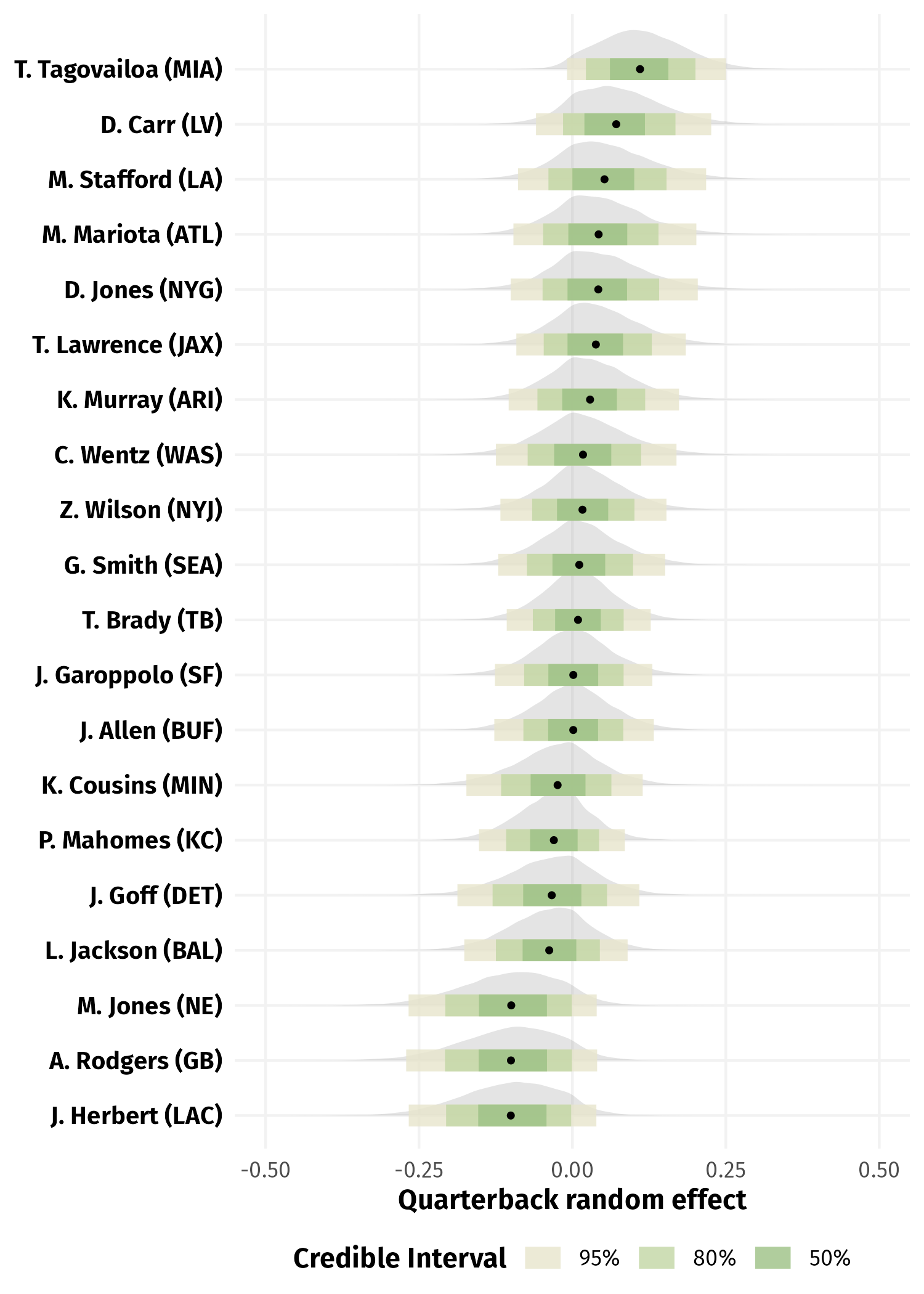} 

}

\caption{Posterior distributions of the quarterback random effect $b_q$ when modeling the mean snap timing. Results shown here are for NFL quarterbacks with at least 50 pass attempts on plays with receivers in motion at snap and running a route. For each player, the posterior mean point estimate and corresponding credible intervals are depicted.}\label{fig:fig_qb_mean_posterior_distributions}
\end{figure}

Here, we investigate the random effects in the mean parameter model for
the two groups on offense: quarterbacks and motion players, resulting
from fitting the multilevel model for snap timing described in Section
\ref{sec:model}. In Figure
\ref{fig:fig_qb_mean_posterior_distributions}, we visualize the
posterior distributions for the random effect \(b_q\) (sorted by
posterior mean) for the same subset of quarterbacks as before (i.e.,
those attempting at least 50 passes across the considered plays). Among
these quarterbacks, Tua Tagovailoa tends to have longer snap timing on
average, whereas Justin Herbert and Aaron Rodgers appear to be quicker
with controlling the snap after motion from their teammates.

\begin{figure}[ptb]

{\centering \includegraphics{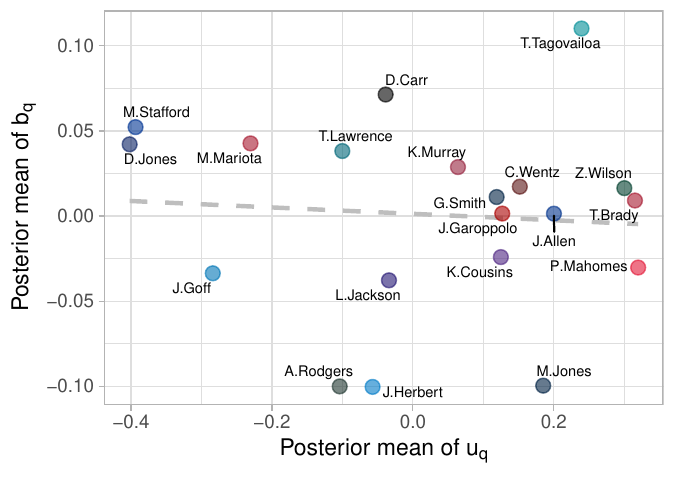} 

}

\caption{Relationship between the posterior means of the quarterback random effects $b_q$ and $u_q$ when modeling the mean and shape snap timing, respectively $(r= -0.07)$. Results shown here are for NFL quarterbacks with at least 50 pass attempts on plays with receivers in motion at snap and running a route.}\label{fig:fig_corr_mean_shape}
\end{figure}

We also display the joint distribution for the posterior mean estimates
of the QB mean and shape random effects (\(b_q\) and \(u_q\),
respectively) in Figure \ref{fig:fig_corr_mean_shape}. As we can see,
the scatterplot reveals no inherent relationship between the average and
variability in snap timing. Additionally, the QB ranking at the mean
level in Figure \ref{fig:fig_qb_mean_posterior_distributions} is also
different from the QB snap timing leaderboard in Section
\ref{sec:leaderboard}. Hence, a quarterback can be consistent or
variable with managing snap timing, and this does not necessarily depend
on the duration of the snap timing itself on average.

\begin{figure}[ptb]

{\centering \includegraphics[width=0.75\linewidth]{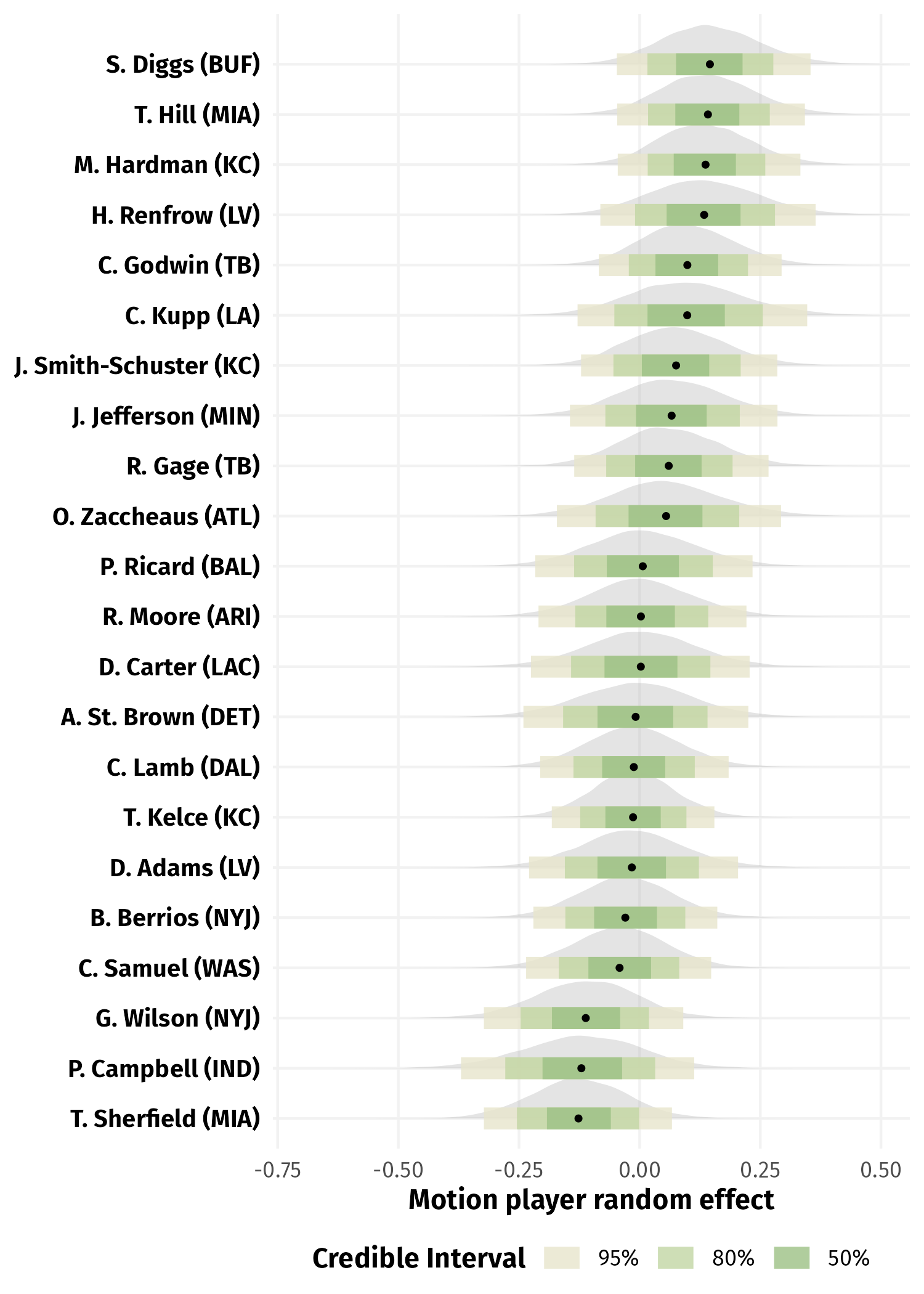} 

}

\caption{Posterior distributions of the motion player random effect $b_m$ when modeling the mean snap timing. Results shown here are for NFL receivers with at least 20 times going in motion on plays with receivers in motion at snap and running a route. For each player, the posterior mean point estimate and corresponding credible intervals are depicted.}\label{fig:fig_receiver_posterior_distributions}
\end{figure}

Next, Figure \ref{fig:fig_receiver_posterior_distributions} shows the
posterior distributions for the motion player random effect \(b_m\) for
those in motion 20 times or more across the motion plays in our final
sample. We notice that Stefon Diggs and Tyreek Hill are the top two
players in terms of their posterior mean values. These are two of the
most elite wide receivers in 2022, with Tyreek Hill especially known for
his lightning-fast speed on the football field and was part of the
electric 2022 Miami Dolphins offense. Thus, at first glance, one could
suspect that these estimates are indicative of receiver speed and
agility to some degree.

However, in each of Figures
\ref{fig:fig_qb_mean_posterior_distributions} and
\ref{fig:fig_receiver_posterior_distributions}, we notice an overlap
between the 95\% credible intervals for the top and bottom players.
Therefore, unlike the variability in QB snap timing, our estimates when
modeling the mean \(\mu\) for both quarterbacks and motion players do
not appear to differentiate between players. We also observe relatively
wide credible intervals for all players, and none of the posterior
distributions are entirely above or below zero. Hence, there is
considerable uncertainty in our estimates for both offensive groups,
which once again highlights the issue of having only a limited sample of
data.

\end{document}